\documentclass[11pt]{article}
\usepackage{moriond,epsfig}
\usepackage{color}

\newcommand\ignore[1]{}
\def\be{\begin{equation}}
\def\ee{\end{equation}}
\def\bea{\begin{eqnarray}}
\def\eea{\end{eqnarray}}

\begin{document}
\vspace*{4cm}
\title{The AdS Graviton/Pomeron Description of  Deep Inelastic Scattering at Small x }

\author{ Richard C. Brower~\footnote{speaker}}

\address{Department of Physics, Boston University, Boston MA 02215}

\author{Marko Djuri\'c}

\address{Centro de F\'isica do Porto, Universidade do Porto, 4169-007 Porto, Portugal}

\author{Ina Sar\v{c}evi\'c}

\address{Physics Department, University of Arizona, Tucson AZ 85721, Department of Astronomy and Steward Observatory, University of Arizona, Tucson, AZ 85721}

\author{Chung-I Tan}

\address{Physics Department, Brown University, Providence RI 02912}

\maketitle

\abstracts{ In the holographic or AdS/CFT  dual to QCD, the Pomeron is identified with a Reggeized Graviton in $AdS_5$~\cite{Brower:2006ea,Brower:2007qh,Brower:2007xg}. We emphasize the importance of confinement, which in this context corresponds to a deformation of $AdS_5$ geometry in the IR.   The holographic  Pomeron  provides
a very good  fit~\cite{Brower:2010wf}  to the  combined data from HERA  for  Deep Inelastic Scattering at small $x$, lending new confidence to this
AdS dual approach to  high energy diffractive scattering.
 }

\paragraph{Pomeron-Graviton Duality:}\label{sec:pomeron}

Traditionally, Deep Inelastic Scattering (DIS)  at small-x, at least for $Q^2$ large, has been modeled  using perturbative QCD. At   small to moderate $Q^2$, confinement  should be taken into account but  it is often ignored, or incorporated in an ad hoc fashion.  Here we use a formulation based on the AdS/CFT correspondence at strong coupling, which has the advantage of a unified soft and hard diffractive
mechanism.  We show, in particular, that the $Q^2$ dependence for the ``effective Pomeron intercept", $\alpha_{eff}=1+\epsilon_{eff}(Q^2)$,  observed at HERA,  can be understood  in terms of diffusion in  $AdS_3$ in the  holographic approach. In this analysis, the bare BPST Pomeron intercept is taken to be $j_0=1.22$.

In lowest order in weak 't Hooft coupling for QCD, a bare Pomeron was first identified by Low and Nussinov as a two gluon exchange corresponding to a Regge cut in the $J$-plane at $j_0 = 1$.   Going beyond the leading order, Balitsky, Fadin, Kuraev and Lipatov (BFKL) summed all the diagrams for two gluon exchange to first order in $\lambda = g^2 N_c$ and {\em all} orders in $(g^2 N_c \log s)^n$, thus giving rise to the so-called BFKL Pomeron. The position of this $J$-plane cut is at $j_0 = 1+ \log (2) g^2 N_c /\pi^2$, recovering the Low-Nussinov result in the $\lambda\rightarrow 0$ limit. In a holographic approach to diffractive scattering~\cite{Brower:2006ea,Brower:2007qh,Brower:2007xg,Cornalba:2006xm}, the weak coupling Pomeron is replaced by the ``Regge graviton'' in AdS space, as formulated by Brower, Polchinski, Strassler and Tan (BPST)~\cite{Brower:2006ea,Brower:2007xg} which has both hard components due to near conformality in the UV and soft Regge behavior in the IR.  Corrections to the  strong coupling lower the intercept from $j=2$ to
\be
j_0 = 2 - 2 /\sqrt{g^2N_c}    \; .
\label{eq:BPST-intercept}
\ee
\begin{figure}
\begin{center}
\hskip .25cm 
\includegraphics[height=0.325 \textwidth,width=.425\textwidth]{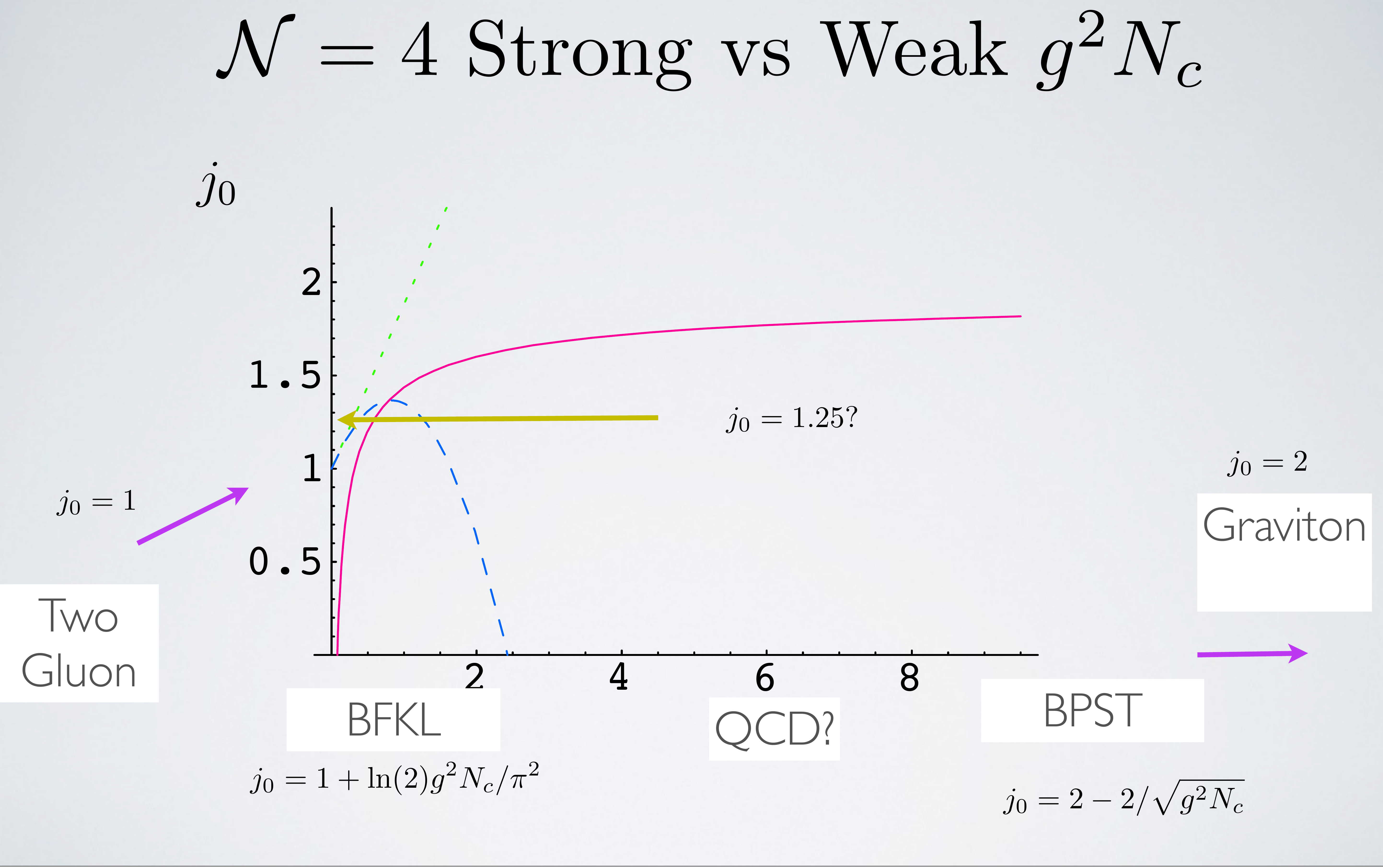}
\includegraphics[height=0.35 \textwidth,width=.45\textwidth]{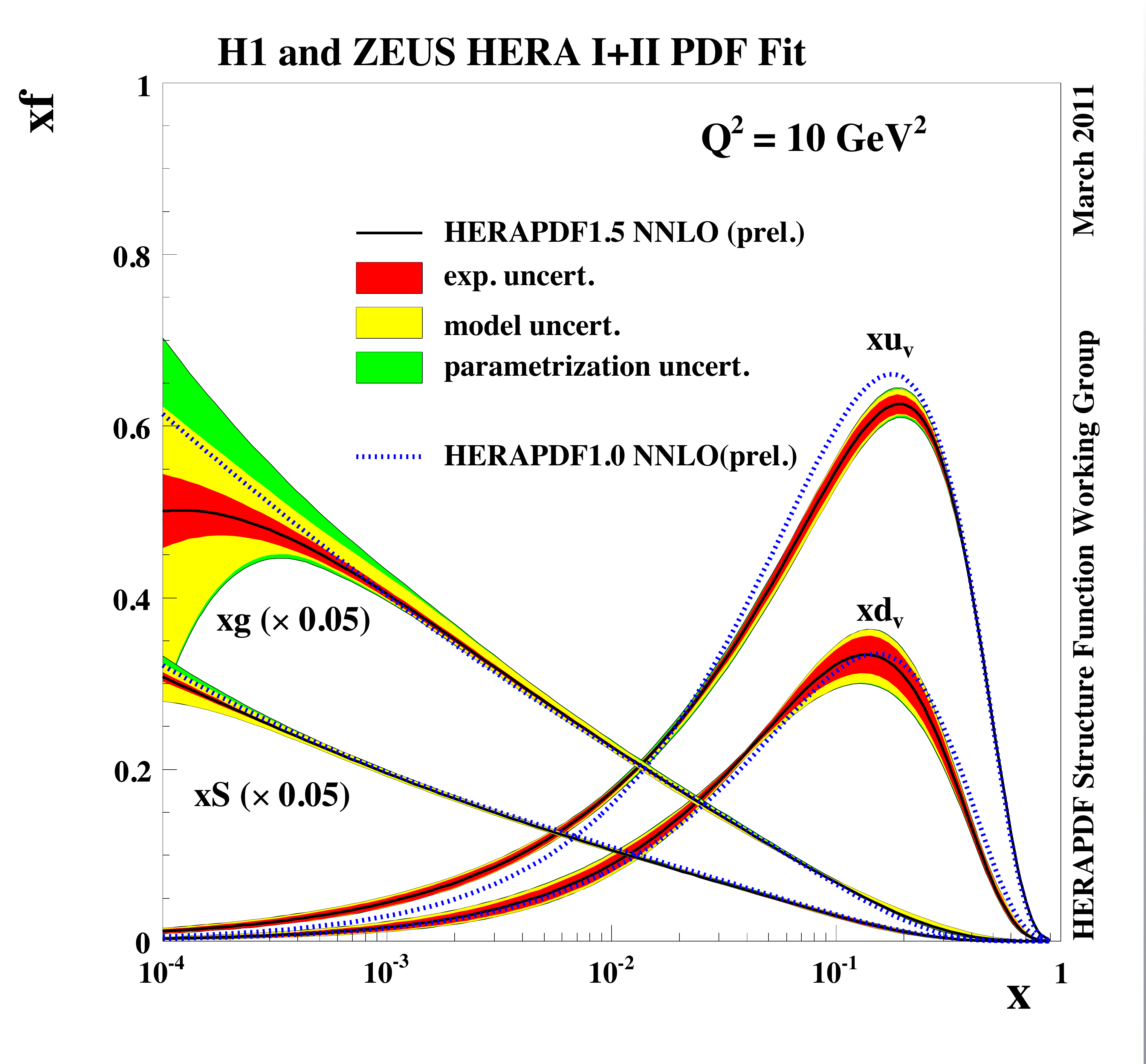}
\end{center}
\caption{ On the left,  intercept $j_0$ in ${\cal N}=4$ YM shown as a function of 't Hooft coupling $\lambda$ for the BPST Pomeron (solid red) and for BFKL (dotted and dashed to first and second order in $\lambda$ respectively). On the right, a typical partonic fit to HERA DIS data demonstrating the dominance for gluon dynamics at small $x$.  }
\label{fig:effective}
\end{figure}
 In Fig.~\ref{fig:effective},  we compare this  with the weak coupling BFKL intercept to second order.  A typical phenomenological estimates for this parameter  for
QCD is about $j_0 \simeq 1.25$,  which suggests that the physics of diffractive scattering is in the cross over region between
strong and weak coupling.  A corresponding treatment for Odderons has also been carried out~\cite{Brower:2008cy}. 
 We also show in Fig.~\ref{fig:effective} the dominance of gluons, in a conventional partonic approach, thus further justifying the large $N_c$ approximation, where  quark constituents are suppressed.

\paragraph{Holographic Treatment to Deep Inelastic Scattering:}\label{sec:DIS}

We  make use of the fact that the DIS cross section can be related to the Pomeron exchange amplitude via the optical theorem, $\sigma=s^{-1}{\rm Im} A(s,t=0)$. In the holographic approach, the impact parameter space $(b_\perp, z)$ is 3 dimensional, where  $z \ge 0$ is the warped radial 5th dimension.  Conformal dilatations (
$\log z \rightarrow \log z + \mbox{const}$) take one from the UV boundary at $z = 0$ deep into the IR  $z = \mbox{large}$.    The near forward elastic amplitude  takes the 
eikonal form,
\begin{equation}
A(s,t)=2i s\int d^{2}b \; e^{i \vec q \cdot \vec b} \int dzdz'\; P_{13}(z)P_{24}(z')\big\{ 1- e^{i \chi(s,b,z,z')}\big\}  \; .\label{eq:A}
\end{equation}
where  $t=-q^2_{\perp}$ and the eikonal function,
$\chi$,
is related to a BPST Pomeron kernel in a transverse $AdS_3$ representation, $\mathcal{K}({s},b,z,z') $,  by $\chi( s,b,z,z')=\frac{g_{0}^{2}}{2{s}}(\frac{R^{2}}{zz'})^2\mathcal{K}({s},b,z,z') $. 
AdS/CFT correspondence gives expressions for all structure functions $F_i$; we focus here on  $F_2(x,Q^2) =\frac{Q^2}{\pi^2\alpha_{em}}(\sigma_T(\gamma^*p)+\sigma_L(\gamma^*p))$.
An  important unifying features for the holographic map is factorization in the AdS space.  For hadron-hadron scattering, $P_{ij}(z)= \sqrt{-g(z)} (z/R)^2 \phi_i(z) \phi_j(z) $  involves a product of two external normalizable wave functions for the projectile and the target respectively.  For DIS, states 1 and 3 are replaced by currents, and we can simply replace $P_{13}$ by product of the  appropriate unnormalized wave-functions.  In the conformal limit, $P_{13}$ was calculated in \cite{Polchinski:2002jw} in terms of Bessel functions, so that, to obtain $F_2$, we simply  replace in (\ref{eq:A}), 
\begin{equation}
P_{13}(z)\rightarrow P_{13}(z,Q^2)=\frac{1}{z} (Qz)^{2}(K_{0}^{2}(Qz)+K_{1}^{2}(Qz))
\end{equation}
When expanded to first order in $\chi$, Eq. (\ref{eq:A}) provides the contribution from exchanging a single Pomeron. In the conformal limit, a simple expression can be found. Confinement can next be introduced, eg., via a hardwall model $z > z_{cut-off}$. The effect of saturation can next be included  via the full transverse $AdS_3$ eikonal representation (\ref{eq:A}). 

\paragraph{Pomeron Kernel:}

The leading order BFKL Pomeron has  remarkable properties. It  enters into the first term in the large $N_c$ expansion with zero beta function.  Thus it is in effect the 
weak coupling cylinder graph for the  Pomeron for a large $N_c$  conformal theory, the same approximations used in the
AdS/CFT approach albeit at strong coupling. Remarkable BFKL integrability properties allows one to treat the BFKL  kernel 
as the solution to  an  $SL(2,\mathcal{C})$ conformal spin chain. Going to strong coupling, the  two gluon  exchange  evolves into a close string
of infinitely many tightly bound  gluons but the same underlying symmetry persists, referred to as  M\"obius invariance  in string theory or the
isometries of the transverse $AdS_3$ impact parameter geometry.  The position of the $j$-plane cut moves
from  $j_0 = 1+ \log (2) g^2 N_c/\pi^2$  up to $j_0 = 2- 2/\sqrt{g^2 N_c} $ and the kernel
obeys a Schr\"odinger equation on $AdS_3$ space for the Lorentz boost operators $M_{+-}$ ,
\begin{equation}
\left[ (-\partial_u^2 - te^{-2u})/2+\sqrt{\lambda}(j-j_0) \right]G_j(t,z,z')=\delta(u-u'),\label{adseq}
\end{equation}
with $z=e^{-u}.$   In the conformal limit,   $G_j(t,z,z')= \int dq\; q \; J_{\tilde \Delta(j)}(zq) J_{\tilde\Delta(j)}(qz')/(q^2-t)$, $\tilde\Delta(j)^2 = 2\lambda (j-j_0)$, and the Pomeron kernel is  obtained via an inverse Mellin transform.  At $t=0$ the solution  for the imaginary part of the Pomeron kernel exhibits diffusion
\begin{equation}
{\rm Im}\; \mathcal{K}(s, t=0, z,z') \sim \frac{s^{\textstyle j_0}}{\sqrt{\pi\mathcal{D}\log s}}\; e^{\textstyle -(\log z-\log z')^2/\mathcal{D}\log s},\label{strongkernel}
\end{equation}
in the "size" parameter $\log z$  for the exchanged closed string, analogous to the BFKL kernel  at weak coupling, with  diffusing taking place in  $\log(k_\perp)$,  the virtuality of the off shell gluon dipole. 
The diffusion constant  takes on  $\mathcal{D} = 2/\sqrt{g^2N_c}$ at strong coupling compared to $\mathcal{D}  = 7 \zeta(3) g^2 N_c/2 \pi^2 $ in weak coupling.
The close analogy between the weak and strong coupling Pomeron
suggests the development of a hybrid phenomenology leveraging plausible interpolations between the two extremes.

\paragraph{Fit to HERA Data}

Both of these integrals, $z$ and $z'$ in (\ref{eq:A}),   remain sharply peaked, the first around $z\sim 1/Q$ and the second around the inverse proton mass, $z'\equiv 1/Q'\sim 1/m_p$. We  approximate both of them by delta functions.   Under such an ``ultra-local" approximation, all structure functions take on very simple form,  e.g, 
\begin{equation}
F_{2}(x,Q^2) =\frac{g_{0}^2 }{8 \pi^{2}\lambda} \frac{Q}{Q'}\frac{e^{\textstyle (j_0 -1)\;\tau} }{\sqrt{\pi {\cal D} \tau }} \;e^{\textstyle -  (\log Q-\log Q')^2/ {\cal D} \tau}     + {\rm Confining \;\; Images}.\label{eq:f2conformalb}
\end{equation}
with  diffusion time given more precisely as $\tau =  \log ( s/QQ'\sqrt{\lambda}) =  \log (1/x) - \log (\sqrt{\lambda}  Q'/Q)$.  Here the first term is conformal and, for hardwall, the confining effect can be expressed in terms of image charges \cite{Brower:2010wf}. It is important to note  that taking the $s\rightarrow \infty$ limit, the amplitude corresponding to (\ref{eq:f2conformalb})  grows asymptotically as $(1/x)^{j_0} \sim s^{j_0}$, thus violating the Froissart unitarity bound at very high energies. The eikonal approximation in $AdS$ space \cite{Brower:2007qh,Cornalba:2006xm,Albacete:2008ze} plays
the role of   implementing ``saturation"  to restore  unitary via multi-Pomeron shadowing.

\begin{figure}
\begin{center}
\includegraphics[height=0.35 \textwidth,width=.45\textwidth]{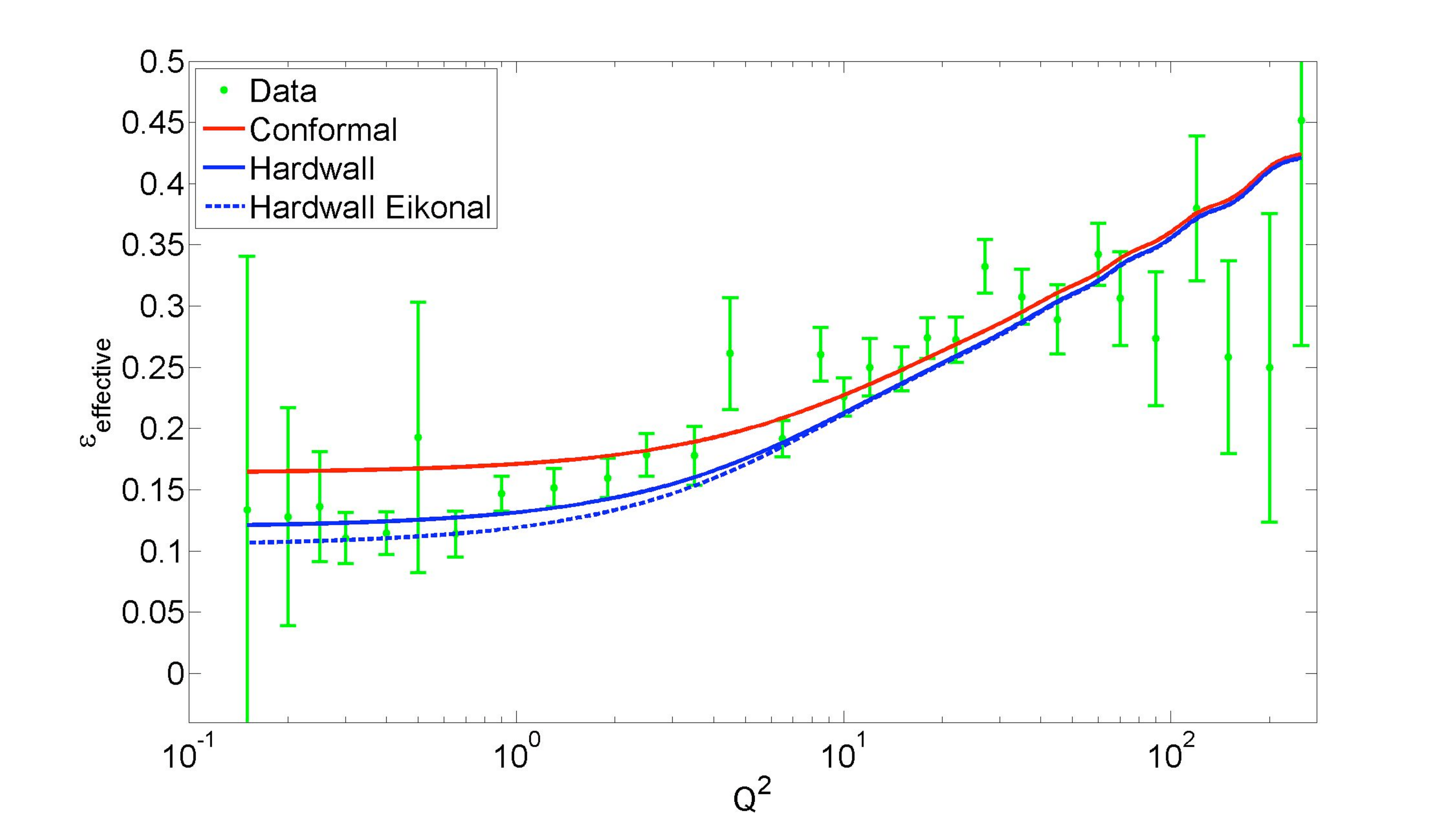}
\includegraphics[height=0.35 \textwidth,width=.45\textwidth]{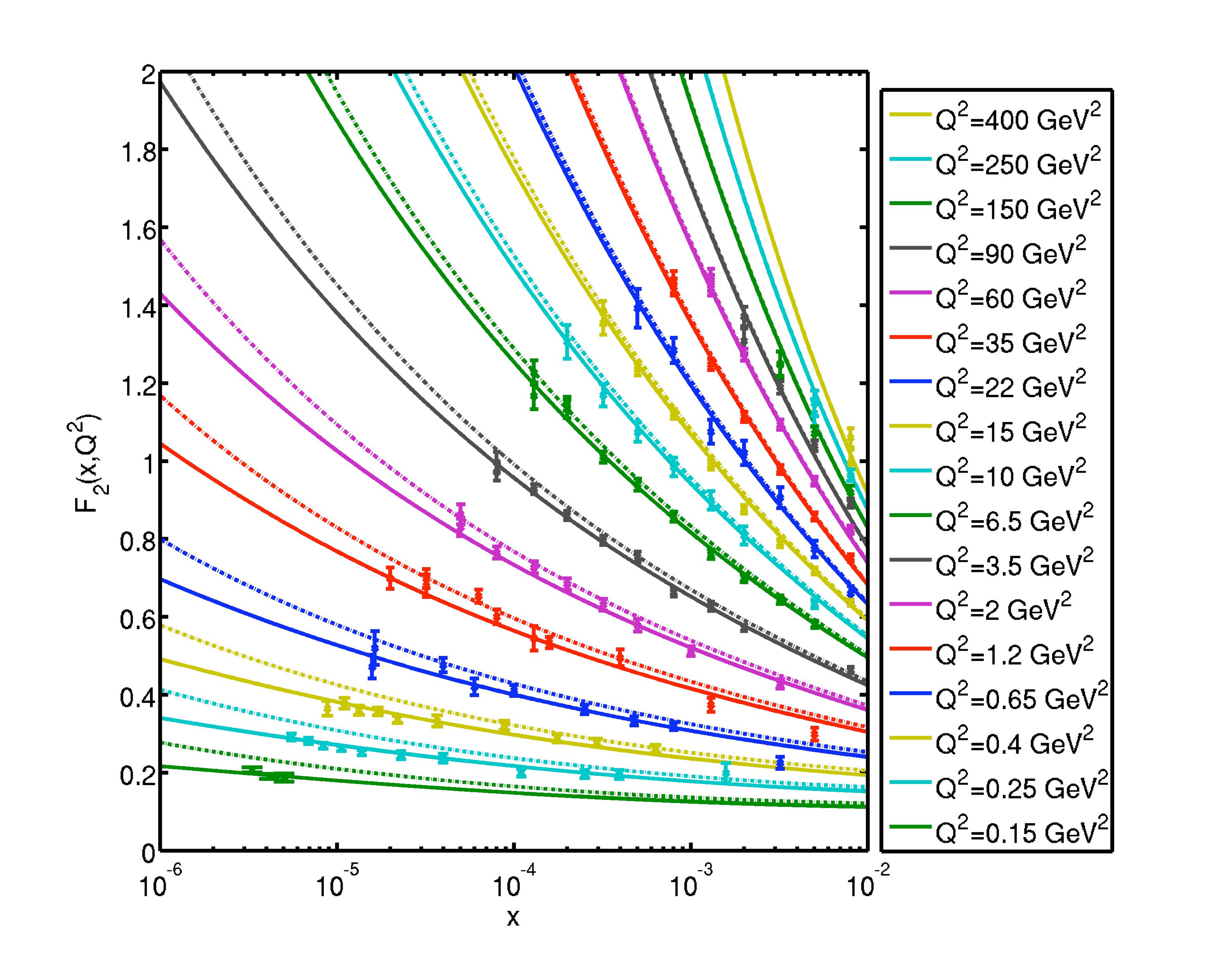}
\end{center}
\caption{In the left, with the BPST Pomeron intercept at 1.22, $Q^2$ dependence for ``effective intercept" is shown for conformal, hardwall and hardwall eikonal model. In the right, a more  detailed fit is presented contrasting the fits to HERA data at small x by a single hardwall Pomeron vs hardwall eikonal respectively.}
\label{fig:HERALHC}
\end{figure}

We have shown various comparisons of our results~\cite{Brower:2010wf} to the data from the combined H1 and ZEUS experiments at HERALD~\cite{:2009wt} in Fig.~\ref{fig:HERALHC}. Both the conformal, the hard-wall model as well as the eikonalized hard-wall model can fit the data reasonably well. This can best be seen in the left figure which exhibits the $Q^2$ dependence of an effective Pomeron intercept. This can be understood  as a consequence of diffusion. However, it is important to observe that  the hard-wall model provides a much better fit than the conformal result  for $Q^2$ less than $2\sim 3 $ $GeV^2$. 
The best fit to data is obtained using the hard-wall eikonal model, with a $\chi^2 = 1.04$.  This is clearly shown by the  figure to the right, where we present a comparison of the relative importance of confinement versus eikonal at the current energies.
We observe  that the   transition scale $Q_{c}^2(x)$  from conformal to confinement increases with  $1/x$, and it comes before saturation effect becomes important.  For more details, see Ref. ~\cite{Brower:2010wf}

\paragraph{Conclusions:}

We have presented the phenomenological application of the AdS/CFT correspondence to the study of high energy diffractive scattering
for QCD.  Fits to the HERA DIS data at small x demonstrates that the strong coupling BPST Graviton/Pomerons~\cite{Brower:2006ea}  does allow for a
very good description of diffractive DIS with few phenomenological parameters, the principle one being the intercept to the bare Pomeron fit to be $j_0 \simeq   1.22$.
Encouraged by this, we plan to undertake a fuller study of several closely related diffractive process: total and elastic cross sections, DIS, virtual photon production
and double diffraction production of heavy quarks. The goal is that by over constraining the basic AdS building blocks of diffractive
scattering, this framework will give 
a compelling phenomenology prediction for the double diffractive production of the Higgs in the standard model to aid in the analysis of LHC data.

\paragraph*{Acknowledgments:}

The work of RCB was supported by the Department of Energy under contract
DE-FG02-91ER40676, that of IS by the Department of Energy under contracts DEFG02-
04ER41319 and DE-FG02-04ER41298 that of CIT by the Department of
Energy under contract DE-FG02-91ER40688, Task-A, and that of MD by FCT project CERN/FP/116358/2010.

\paragraph*{References:}

\end{document}